\documentclass[twocolumn,showpacs,superscriptaddress,amsmath,aps]{revtex4}
\usepackage{graphicx,color}
\usepackage{bm}
\usepackage[hypertex]{hyperref}

\newcommand{\be}{\begin{equation}}
\newcommand{\ee}{\end{equation}}
\newcommand{\bea}{\begin{eqnarray}}
\newcommand{\eea}{\end{eqnarray}}
\newcommand{\bsube}{\begin{subequations}}
\newcommand{\esube}{\end{subequations}}

\newcommand{\Eq}[1]{Eq.\,(\ref{#1})}
\newcommand{\Eqs}[1]{Eqs.\,(\ref{#1})}

\newcommand{\la}{\langle}
\newcommand{\ra}{\rangle}

\newcommand{\nl}{\nonumber \\}

\newcommand{\bsub}{\begin{subequations}}
\newcommand{\esub}{\end{subequations}}

\begin{document}
%\begin{CJK*}{GBK}{Song}

\title{Qubit state tomography in superconducting circuit
        via weak measurements}

\author{Lupei Qin}
\affiliation{Center for Advanced Quantum Studies and
Department of Physics, Beijing Normal University,
Beijing 100875, China}

\author{Luting Xu}
\affiliation{Center for Advanced Quantum Studies and
Department of Physics, Beijing Normal University,
Beijing 100875, China}

\author{Wei Feng}
\affiliation{Department of Physics, Tianjin University,
Tianjin 300072, China}

\author{Xin-Qi Li}
\email{lixinqi@bnu.edu.cn}
\affiliation{Center for Advanced Quantum Studies and
Department of Physics, Beijing Normal University,
Beijing 100875, China}

\date{\today}

%% \maketitle
\begin{abstract}
The standard method of ``measuring" quantum wavefunction
is the technique of {\it indirect} quantum state tomography.
Owing to conceptual novelty and possible advantages,
an alternative {\it direct} scheme was proposed
and demonstrated recently in quantum optics system.
In this work we present a study on the direct scheme of
measuring qubit state in the circuit QED system,
based on weak measurement and weak value concepts.
To be applied to generic parameter conditions,
our formulation and analysis are carried out
for finite strength weak measurement, and
in particular beyond the bad-cavity and weak-response limits.
The proposed study is accessible to the present
state-of-the-art circuit-QED experiments.
\end{abstract}


\maketitle

\section{Introduction}

{\flushleft
In quantum mechanics}, the state of a single system
(e.g., a single particle) is described by a wavefunction,
which differs drastically from the description in classical mechanics.
It is well known that the wavefunction cannot be
determined via a single shot measurement \cite{WZ82}.
Actually, the wavefunction is not a physical quantity,
but a {\it knowledge} governed by Schr\"odinger equation
and utilized to calculate the real
physical quantities by means of statistical average.
However, with the advent of quantum information science and technology,
experimental manipulation and determination of the wavefunction
have become extremely important.

% (传统的“间接方法”)
In order to determine the wavefunction,
the standard method is based on projective strong measurement
where the wavefunction is fully collapsed, and has been termed
as {\it quantum state tomography} \cite{Ris89,Smi93,Bre97,Kwi99,Hof09}.
In that, one must perform a large set of distinct measurements
on many identical copies of the system and reconstruct the state
that is most compatible with the measurement results.
This tomographic reconstruction of quantum state
is considered ``indirect" determination,
owing to the requirement of post-processing.

% (新的“直接方法”)
An alternative scheme is the so-called
``direct" state determination \cite{Lun11,Lun12,Boy13,Boy14},
which may have potential applications
in quantum information and quantum metrology.
This method is based on the idea of
sequentially measuring two complementary variables of the system
\cite{Aha88,Ste89,Aha90,Rit91,Wis02,Pop04,
Joh04,Wis05,Kwi08,Jor09,Koc11,Ste11}.
The first measurement is weak, and the second one is strong (projective).
The weak measurement (each single one) gets minor information,
has gentle disturbance, and does not collapse the state.
The second projective measurement plays a role of post-selection.
Under this sort of joint measurements,
the real and imaginary components of the wavefunction
will appear directly on the measurement apparatus,
in terms of a shift of the pointer's position
by amount of the weak value (WV) introduced by Aharonov,
Albert and Vaidman (AAV) nearly 30 years ago,
given by \cite{Aha88,Aha90}
\bea\label{AAV-WV}
A_w = \frac{\langle \psi_f|\hat{A}| \psi_i\rangle}
{\langle \psi_f|\psi_i\rangle}  \,,
\eea
where $|\psi_i\rangle$ and $|\psi_f\rangle$ are, respectively,
in the context of state tomography,
the state to be determined and the one for post-selection.
$\hat{A}$ is the weakly observed quantity.
%%  (直接性)
The main advantage of this method is that
it is free from complicated sets
of measurements and computations,
manifesting a ``directness" of no need
of post-processing the average raw signal.
%% exp 情况
For instance, applying this novel approach,
recent experiments have been carried out for the
{\it direct} measurement of photon's transverse wavefunction
(a task not previously realized by any method) \cite{Lun11},
and for a full characterization of polarization states
of light via direct measurement \cite{Boy13}.

In this scheme, weak measurement is at the heart.
On the other hand,
the superconducting circuit quantum electrodynamics (cQED) system
is currently an important platform for performing
quantum weak measurement and control studies
\cite{Bla04,Wall04,Sid12,Dev13,Sid13,Sid15}.
In this work, we present an analysis on the possible
direct measurement of qubit state in this system.
In order to be applied to generic parameter conditions,
our study will be put on finite strength
of weak measurement \cite{Jor08,Li15}.
This goes beyond the usual limit of vanishing strength,
thus results in a generalized pre- and post-selection (PPS) average,
rather than the original AAV WV, as the pointer's shift of apparatus.
To extract the AAV WV from the PPS average (raw signal),
we propose to apply the analytic formula derived
for the homodyne measurement in circuit QED \cite{Li15}.
By varying the local oscillator's (LO) phase,
one can easily extract the complex weak value
and determine the complex wavefunction,
applying a simple iterative algorithm.
For the first time, we also obtain analytic result
for the PPS average beyond the bad-cavity and weak-response limits,
and demonstrate how to reliably determine the qubit state in this regime.

\section{Methods}

\subsection{Measurement Current and Rates}
% {\flushleft \it Measurement Current and Rates}.---
%%
{\flushleft The solid-state circuit QED (cQED) system} is originally
described by the well-known Jaynes-Cummings (J-C) model \cite{Bla04,Wall04}.
In dispersive regime, i.e., the detuning $\Delta$
between the cavity resonance frequency and the qubit energy
being much larger than the J-C coupling strength $g$,
the interaction Hamiltonian is reduced as \cite{Bla04,Wall04}
\bea
H_{\rm int}= \chi a^{\dagger}a\sigma_z  \,,
\eea
where $\chi=g^2/\Delta$ is the dispersive coupling,
$a^{\dagger}$ and $a$ are the
creation and annihilation operators of the cavity photon,
and $\sigma_z$ is the Pauli operator for the qubit.
This type of interaction allows for a homodyne measurement
with output current \cite{Gam08}
\bea\label{I-t}
I(t)=-\sqrt{\Gamma_{ci}(t)}\langle \sigma_{z}\rangle+\xi(t),
\eea
where $\xi(t)$ is a Gaussian white noise
originated from fundamental quantum-jumps during the measurement.
This expression of current was obtained
in the absence of qubit rotation
and from eliminating the cavity
degrees of freedom (the so-called polaron transformation) \cite{Gam08}.
In a bit more detail, $\Gamma_{ci}(t)$ is
the measurement information gain rate given by \cite{Gam08}
\bea
\Gamma_{ci}(t) = \kappa|\beta(t)|^2\cos^2(\varphi-\theta_\beta) \, ,
\eea
where $\varphi$ is the local oscillator's (LO)
phase in the homodyne measurement,
$\kappa$ is the leaky rate of the cavity photons,
and $\beta(t)=\alpha_2(t)-\alpha_1(t)
\equiv |\beta(t)|e^{i\theta_{\beta}}$
with $\alpha_1(t)$ and $\alpha_2(t)$ being the cavity fields
associated with the qubit states $|1\ra$ and $|2\ra$, respectively.

In addition to the information gain rate $\Gamma_{ci}$,
there exists as well a {\it no-information}
back-action rate which reads \cite{Gam08}
\bea
\Gamma_{ba}(t) = \kappa|\beta(t)|^2\sin^2(\varphi-\theta_\beta)\,.
\eea
This corresponds to the ``realistic" backaction rate
discussed in Ref.\ \cite{Kor11},
while $\Gamma_{ci}$ is named ``spooky" backaction rate.
To understand the physical meaning, let us consider
the stochastic evolution of
qubit state $c_1(t)|1\ra+c_2(t)|2\ra$,
conditioned on measurement records in a single realization.
The rate $\Gamma_{ci}$ appearing \Eq{I-t} is
associated with distinguishing the qubit state (information gain),
which causes change of the probability amplitudes
of $|1\ra$ and $|2\ra$. As opposed to this,
the rate $\Gamma_{ba}$ is only related to phase fluctuation
between $|1\ra$ and $|2\ra$ (no probability change).
The sum of $\Gamma_{ci}$ and $\Gamma_{ba}$,
$\Gamma_{m}= \Gamma_{ci}+\Gamma_{ba}$,
gives the total measurement rate.
Differing somehow from $\Gamma_m$,
the overall decoherence rate is given by \cite{Gam08}
\bea
\Gamma_d(t)=4\chi\mathrm{Im}[\alpha_1(t)\alpha^*_2(t)] \,,
\eea
as a result from tracing the cavity degrees of freedom
from the whole entangled qubit-cavity state.
An interesting point is that $\Gamma_m$
is not necessarily equal to $\Gamma_d$,
owing to ceratin ``information loss".
Only for ideal (quantum limited) measurement,
$\Gamma_{m}=\Gamma_d$
and the single quantum trajectory is
a quantum mechanically pure state.

\subsection{Quantum Bayesian Rule}
%{\it Quantum Bayesian Rule}.---
%%
{\flushleft
Conditioned on the output currents}, \Eq{I-t},
one can faithfully keep track of
the stochastic evolution of the qubit state.
In order to get analytic expression for the PPS average,
rather than the quantum trajectory equation,
we apply alternatively the quantum Bayesian rule (BR)
\cite{Kor99,Kor11,Li14,Li16}.
Using the output currents during $(0,t_m)$,
we update first the diagonal elements $\rho_{jj}$ ($j=1,2$)
of the qubit state (density matrix) \cite{Li14,Li16}
\bea\label{BR-cQED-b}
\rho_{jj}(t_m) = \rho_{jj}(0) \, P_j (t_m)/ {\cal N}(t_m) \,,
\eea
where
${\cal N}(t_m)=\sum_{j=1,2}\rho_{jj}(0) \, P_j (t_m)$.
This result simply follows the standard Bayes formula,
and the {\it functional} distribution of currents reads \cite{Li16}
\bea\label{P12}
P_{1(2)}(t_m)
= \exp\left\{-\la [I(t)-\bar{I}_{1(2)}(t)]^2 \ra_{t_m} / (2V) \right\} \,,
\eea
where
$\bar{I}_{1(2)}(t)=\mp\sqrt{\Gamma_{ci}(t)}$
and $\la \bullet \ra_{t_m}=(t_m)^{-1}\int^{t_m}_{0} dt (\bullet)$,
and $V=1/t_m$ characterizes the distribution variance.
The result of \Eq{P12} differs from our usual knowledge.
From quite general consideration (central limit theorem),
corresponding to $|1\ra$ and $|2\ra$,
the averaged stochastic current $I_m$,
$I_m=(t_m)^{-1}\int^{t_m}_{0} dt I(t)$,
should respectively center at
$\bar{I}_{1(2)}=\mp (t_m)^{-1}\int^{t_m}_{0} dt\,
\sqrt{\Gamma_{ci}(t)}$
and satisfies the standard Gaussian distribution:
\bea\label{G-P12}
P_{1(2)}(t_m)  = (2\pi V)^{-1/2}
\exp\left[-(I_m-\bar{I}_{1(2)})^2  / (2V) \right]  \,.
\eea
In Ref.\ \cite{Li16}, we have demonstrated that
this ``standard" result is valid only for
time-independent rate $\Gamma_{ci}$, in \Eq{I-t}.

Secondly, we update the off-diagonal elements as follows:
% \begin{subequations}\label{BR-cQED}
\bea\label{BR-cQED-a}
&& \rho_{12}(t_m) = \rho_{12}(0)
   \left[\sqrt{P_1(t_m)P_2(t_m)}/{\cal N}(t_m) \right]  \nl
&& ~~~~~~ \times D(t_m) \,
\exp\left\{-i[\Phi_1(t_m)+ \Phi_2(t_m)]\right\}  \,.
\eea
Compared to the original simple BR \cite{Kor99},
a couple of correction factors appear in this result,
specifically, given by \cite{Li14,Li16}
\begin{subequations}\label{BR-factors}
\bea \label{factor-D}
D(t_m)&=& \exp \left\{-\int_0^{t_m}dt [\Gamma_d(t)
-\Gamma_m(t)]/2 \right\}  ,
\eea  %   \\
\bea \label{factor-1}
\Phi_1(t_m)&=&\int_0^{t_m} dt \, \widetilde{\Omega}_q(t)   \,, % \\
\eea
\bea\label{factor-2}
\Phi_2(t_m)&=& -\int_0^{t_m} dt \sqrt{\Gamma_{ba}(t)}\,I(t) \,.
\eea
\end{subequations}
Here we have introduced
$\widetilde{\Omega}_q(t)=\omega_{q}+\chi+\mathrm{B}(t)$,
i.e., the bare qubit energy $\omega_{q}$ is renormalized
by the dispersive shift $\chi$ and ac Stark effect induced shift
$\mathrm{B}(t)=2\chi{\rm Re}[\alpha_1(t)\alpha^*_2(t)]$.
Briefly speaking,
the purity degradation factor $D(t_m)$ is
owing to non-ideality (information loss) in the measurement,
while the two phase factors
$e^{-i\Phi_1(t_m)}$ and $e^{-i\Phi_2(t_m)}$
are resulted, respectively, from the dynamic ac-Stark effect
and the no-information ``realistic" back-action.

\subsection{PPS average under
bad-cavity and weak-response limits}
%{\it Bad-cavity and weak-response limits}.---
%%
{\flushleft In experiments}
the cQED system is usually prepared in the
bad-cavity and weak-response limits. % \cite{Sid12,Dev13,Sid13,Kor11}.
In this case, the cavity-field evolves to stationary state
on timescale much shorter than the measurement time.
One can thus carry out the ac-Stark shift and all the rates
using the {\it stationary} coherent-state fields of the cavity,
$\bar{\alpha}_1$ and $\bar{\alpha}_2$, which read \cite{Li14}
\bea\label{coh-state}
\bar{\alpha}_{1(2)}=-i\epsilon_m/[-i(\Delta_r\pm\chi)+\kappa/2],
\eea
where $\Delta_r=\omega_m-\omega_r$ is the offset
of the measurement and cavity frequencies.
For instance, in the bad-cavity and weak-response limits,
we obtain the stationary $B(t)$ as $B\simeq 2\chi\bar{n}$,
where $\bar{n}=|\bar{\alpha}|^2$
and $\bar{\alpha}= -i\epsilon_m/(\frac{\kappa}{2})$,
which recovers the standard ac-Stark shift.
Also, a resonant drive ($\omega_m=\omega_r$)
results in $\theta_{\beta}=0$.

Now let us consider the weak value for finite strength measurement.
For simplicity, we denote the measurement result as
$x\equiv I_m=(t_m)^{-1}\int^{t_m}_{0}dt I(t)$,
and $\bar{x}_j\equiv \bar{I}_j=(-1)^j\sqrt{\Gamma_{ci}}$.
Under the same spirit of the AAV WV
for infinitesimal strength of measurement,
we employ the following PPS average as a definition
for the WV associated with finite strength measurement
\cite{Wis02,Jor08,Li15}
\bea\label{WV-1}
_{f}\la x \ra_{i} = \frac{\int dx\, x P_{\psi_i}(x) P_x(\psi_f)}
  {\int dx\, P_{\psi_i}(x) P_x(\psi_f)}  \,.
\eea
$P_{\psi_i}(x)$
is the distribution probability of the measurement outcomes
associated with the pre-selected state $|\psi_i\ra$,
before the post-selection using $|\psi_f\ra$ .
$P_{x}(\psi_f)$ is the post-selection probability given by
$P_{x}(\psi_f)= \la\psi_f|\tilde{\rho}(x)|\psi_f\ra$,
by applying the quantum BR to update the state
from $\rho_i$ to $\tilde{\rho}(x)$,
based on the measurement outcome $x$.
After some algebras , we obtain \cite{Li15}
\bea\label{cQED-WV-3}
_f\la x\ra_i = - \frac
{\epsilon_1 {\rm Re}(\sigma^z_w)
+ \epsilon_2 {\rm Im}(\sigma^z_w)}
{1+{\cal G}\, (|\sigma^z_w|^2-1)} \,,
\eea
where $\epsilon_1 = \sqrt{\Gamma_{ci}}$,
$\epsilon_2 = \sqrt{\Gamma_{ba}}\, e^{-\Gamma_d t_m}$,
and ${\cal G}=(1-e^{-\Gamma_d t_m})/2$.
In this result, the AAV WV is slightly modified as
\bea\label{mod-WV}
\sigma^z_w = \frac{\la\psi_f|\sigma_z|\tilde{\psi}_i\ra}
{\la\psi_f|\tilde{\psi}_i\ra} \,,
\eea
where $|\tilde{\psi}_i\ra$ differs from the initial state
$|\psi_i\ra=c_1|1\ra + c_2|2\ra$ by a phase factor as
$|\tilde{\psi}_i\ra
= c_1 e^{-i \widetilde{\Omega}_q t_m}|1\ra + c_2|2\ra$.

We see that,
by tuning the LO phase $\varphi$ based on \Eq{cQED-WV-3},
one can conveniently measure the real and imaginary parts of
$\tilde{\sigma}^z_w$, from which efficient state-tomography
technique can be developed for finite strength measurement.

\subsection{Beyond bad-cavity and weak-response limits}
%{\it Beyond the bad-cavity and weak-response limits}.---
%%
{\flushleft Following the same definition}
of the PPS average, \Eq{WV-1}, we have
\bea\label{exact-WV-2}
_f \langle x\rangle_{i}
&=& \frac{\int {\cal D}[I(t)]\, x[I(t)] \,
P_{\psi_i}(\{I(t)\}) \, P_{\{I(t)\}}(\psi_f)}
{\int {\cal D}[I(t)] \,
P_{\psi_i}(\{I(t)\}) \, P_{\{I(t)\}}(\psi_f)}   \nl
&\equiv& \frac{M_1}{M_2}   \,,
\eea
where
$x[I(t)] = (t_m)^{-1} \int^{t_{m}}_{0}I(t)dt$,
and the two probability distribution functionals read
\begin{eqnarray}
  P_{\psi_i}(\{I(t)\})
  &=& \rho_{11}(0)P_1(t_m) +\rho_{22}(0) P_2(t_m) \,,  \nonumber \\
  P_{\{I(t)\}}(\psi_f)
  &=& \tilde{\rho}_{11}\rho_{f11}+\tilde{\rho}_{22}\rho_{f2}
  +\tilde{\rho}_{12}\rho_{f21}+\tilde{\rho}_{21}\rho_{f12}  \,.   \nonumber
\end{eqnarray}
Here we have denoted the Bayesian updated state by
$\tilde{\rho}=\tilde{\rho}(x)=\tilde{\rho}(\{I(t)\})$.
The probabilities $P_{1,2}(t_m)$ follow \Eq{P12},
being {\it functionals} of the current record
$\{I(t) \,|\, t\in[0,t_m]\}$.
By means of the Gaussian ``path integral" method,
calculation of \Eq{exact-WV-2} is straightforward.
We obtain
\begin{eqnarray}
  M_{1}
  &=& -\left(\int_0^{t_m}\sqrt{\Gamma_{ci}(t)}dt \right )
  (\rho_{11}\rho_{f11}-\rho_{22}\rho_{f22})\nonumber \\
   && + \left (\int_0^{t_m}\sqrt{\Gamma_{ba}(t)}dt \;\; e^{-\int_0^{t_m}\sqrt{\Gamma_{d}(t)}dt} \right )  \nl
   &&  ~~~ \times \, 2 \, \mathrm{Im}(\rho_{21}\rho_{f12}
   e^{i\int_0^{t_m}\tilde{\Omega}(t)dt})  \,,
\end{eqnarray}
\begin{eqnarray}
  M_2
  &=& \rho_{11}\rho_{f11}+\rho_{22}\rho_{f22}
      + \left(e^{-\int_0^{t_m}\sqrt{\Gamma_{d}(t)}dt} \right) \nl
&& \times \,  2\, \mathrm{Re}(\rho_{21}\rho_{f12}
  e^{i\int_0^{t_m}\tilde{\Omega}(t)dt} )   \,.
\end{eqnarray}
Reorganizing this result further in terms of the AAV WV form,
we find that the same expression as \Eq{cQED-WV-3} can be obtained,
with only modifying the several parameters as
\begin{eqnarray}\label{t-factor}
  \epsilon_1 &=& \int_0^{t_m}\sqrt{\Gamma_{ci}(t)}dt  \,, \nl
  \epsilon_2 &=& \int_0^{t_m}\sqrt{\Gamma_{ba}(t)}dt\,e^{-\int_0^{t_m}
  \sqrt{\Gamma_{d}(t)}dt}  \,,  \nl
   {\cal G} &=& (1-e^{-\int_0^{t_m}\sqrt{\Gamma_{d}(t)}dt})/2  \,.
\end{eqnarray}
And, the AAV WV of \Eq{mod-WV} is now modified
by replacing the initial state $|\psi_i\rangle$
with $|\tilde{\psi}_i\rangle = c_1 e^{-i\Phi_1(t_m)}
|1\rangle + c_2|2\rangle$.
$\Phi_1(t_m)$ is given by \Eq{factor-1}.

\subsection{Numerical Methods}
%{\it Numerical Methods}.---
%%
{\flushleft From \Eq{cQED-WV-3} }
we see that, in the {\it weak limit} of measurement
(linear response regime), we may approximate
the denominator by unity (neglecting the second term).
In this case one can obtain ${\rm Re}(\sigma^z_w)$
and ${\rm Im}(\sigma^z_w)$
from the PPS average of currents by choosing, respectively,
the LO's phase $\varphi=0$ and $\pi/2$.
In more general case (nonlinear response regime),
the full denominator of \Eq{cQED-WV-3}
should be taken into account.
In this case one can extract
${\rm Re}(\sigma^z_w)$
and ${\rm Im}(\sigma^z_w)$
by applying an {\it iterative} algorithm.
That is, first, set trial values for the real and imaginary parts
of the AAV WV through
\bea
{\rm Re}(\sigma^z_w)
&\Leftarrow&
- \left(_{f}\langle x\rangle_{i}/\epsilon_1\right)|_{\varphi=0} \,, \nl
{\rm Im}(\sigma^z_w)
&\Leftarrow&
- \left(_{f}\langle x\rangle_{i}/\epsilon_2\right)|_{\varphi=\pi/2}
\,. \nonumber
\eea
Then, iteratively evaluate \Eq{cQED-WV-3} for several or tens of times,
until convergence is reached.

About the accuracy of the AAV WV extracted,
we find that, by simulating $10^6$ trajectories,
the accuracy 0.5\% can be achieved for $\varphi=0$,
while it decreases to 3\% for $\varphi=\pi/2$.
The reason is that, in the latter case,
the information gain component
(the first term) in \Eq{I-t} vanishes,
thus resulting in stronger fluctuations of the output currents.
In practice, one may choose $\varphi=\pi/4$,
rather than $\pi/2$.
Using \Eq{cQED-WV-3}, ${\rm Re}(\sigma^z_{w})$
and ${\rm Im}(\sigma^z_{w})$ can be easily extracted as well.
For this choice, the same accuracy as for $\varphi=0$ can be achieved.

With the knowledge of ${\rm Re}(\sigma^z_{w})$
and ${\rm Im}(\sigma^z_{w})$, based on \Eq{mod-WV},
one can {\it directly} determine the unknown state,
$|\psi_i\rangle=c_1|1\rangle + c_2|2\rangle$, as follows.
Note that owing to the dynamic ac-Stark effect,
the wavefunction involved in the AAV WV
is actually ``modified" as
$|\tilde{\psi}_i\rangle
= c_1 e^{-i\Phi_1(t_m)} |1\rangle + c_2|2\rangle$.
Up to a normalization factor,
we rewrite this unknown state as
\bea
|\tilde{\psi}_i\rangle
= |1\rangle + \tilde{c}\,|2\rangle   \,,
\eea
where $\tilde{c}= (c_2/c_1)  e^{i\Phi_1}$.
For a given post-selection state
$|\psi_f\rangle=b_1|1\rangle + b_2|2\rangle$,
the AAV WV can be expressed as
\bea\label{WF-1}
\sigma^{z}_{w}=\frac{b^*_1-b^*_2 \tilde{c}}
{b^*_1+b^*_2 \tilde{c}}  \,.
\eea
%On the other hand, from the PPS average of experimental data,
%one can extract the real and imaginary parts
%of $\tilde{\sigma}^{z}_{w}$ using Eq(..).
From this result, we obtain
\bea
\tilde{c}=\left(\frac{1-\sigma^{z}_{w}}
{1+\sigma^{z}_{w}}  \right)
\left(\frac{b_1}{b_2}\right)^* \equiv r e^{i\tilde{\theta}}  \,,
\eea
which fully characterizes the unknown state $|\psi_i\ra$
by noting that
$c_2/c_1=r e^{i(\tilde{\theta}-\Phi_1)}$.

\section{Results}

{\flushleft In the whole simulations},
we consider measurement under resonant drive,
which corresponds to $\Delta_r=\omega_r-\omega_d=0$
where $\omega_r$ and $\omega_d$ are, respectively,
the cavity frequency and the
frequency of the driving microwave.
Viewing that most present experiments are performed
in the bad-cavity and weak response regime,
we restrict our simulations for Figs.\ $1\sim 4$ to such limits.
In terms of an arbitrary system of units, we denote the strength
of the microwave drive as $\epsilon_m=1.0$,
then set $\kappa=8$ and $\chi=0.1$
under the bad-cavity and weak-coupling conditions.
Under this choice, one can estimate
the average photon number in the cavity (in steady state)
as $\bar{n}\simeq 0.006$, from $\bar{n}=|\alpha_0|^2$
and $\alpha_0=-i\varepsilon_m/(\frac{\kappa}{2})$.
This weak field in the cavity, together with the weak
dispersive coupling $\chi$, defines also a regime
of weak response (in the sense of measurement signal
to the qubit state).
Only in Fig.\ 5,
we display results beyond the bad-cavity and
weak response limits by setting $\kappa=2$
and keeping $\varepsilon_m$ and $\kappa$ unchanged,
which results in the average cavity photon number
$\bar{n}$=1.0 in steady state.

% ====================

In the following results,
we denote the unknown (to be determined) state as
$|\psi_i\ra=\cos\frac{\theta_i}{2} |1\ra
+ \sin\frac{\theta_i}{2} e^{-i\phi_i}  |2\ra$,
and ``secretly" assign
$\theta_i=\pi/3$ and $\phi_i=50\widetilde{\Omega}_q$.
For the post-selection state $|\psi_f\ra$, we only alter
the polar angle $\theta$ to illustrate the quality of tomography.
In all cases, we run the polaron-transformed
effective quantum trajectory equation \cite{Li15,Gam08},
to generate $10^6$ PPS trajectories.

\begin{figure}
  \includegraphics[scale=0.80] {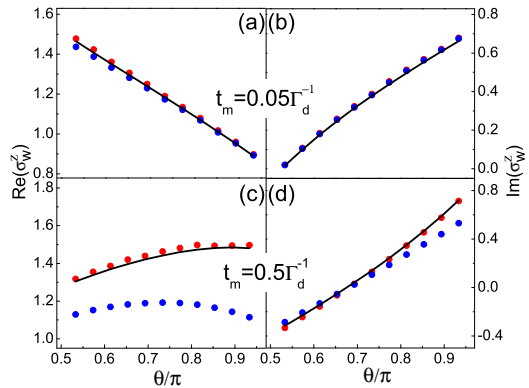}\\
  \caption{
Extracted AAV WVs {\it versus} the post-selection state
(characterized by its polar angle $\theta$).
The correction effect of the second ${\cal G}$ term
in the denominator of \Eq{cQED-WV-3} is illustrated
through its inclusion (red dots) and neglect (blue dots),
in comparison with the ``true" values (solid curves).
Two strengths of measurement are considered:
in (a) and (b), $t_{m}=0.05 \Gamma^{-1}_{d}$;
in (c) and (d), $t_{m}=0.5 \Gamma^{-1}_{d}$.
Parameters:
$\Delta_r=0$, $\epsilon_{m}=1.0$, $\chi=0.1$,
and $\kappa=8.0$.   }
\end{figure}

In Fig.\ 1 we display the extracted AAV WV
against the post-selection state $|\psi_f\ra$.
Our main interest here is the correction effect
of the second ${\cal G}$ term in the denominator of \Eq{cQED-WV-3}.
We thus simulate two strengths of measurement by choosing
the measurement time $t_m=0.05 \Gamma_d^{-1}$
for Fig.\ 1 (a) and (b),
and $t_m=0.5 \Gamma_d^{-1}$ for (c) and (d).
We compare the AAV WVs (the red and blue dots)
extracted from \Eq{cQED-WV-3}
with the ``true" results (solid lines)
calculated using \Eq{mod-WV}
with the ``testing" state $|\psi_i\ra$.
The results of the red dots
are extracted from the full formula of \Eq{cQED-WV-3},
while the blue dots are from neglecting
the second ${\cal G}$ term in the denominator.
We see that for vanishing strength of measurement,
as shown in Fig.\ 1(a) and (b), the effect of the ${\cal G}$ term
is negligible. However, for finite strength of measurement
(Fig.\ 1(c) and (d)),
one must take into account the ${\cal G}$ term.

\begin{figure}
  \includegraphics[scale=0.65] {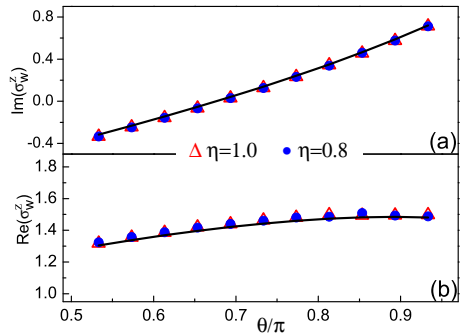}\\
  \caption{
AAV WVs extracted from the PPS averages
under ideal ($\eta=1$) and non-ideal ($\eta=0.8$) measurements,
in comparison with the ``true" values (solid curves).
Parameters:
$\Delta_r=0$, $\epsilon_{m}=1.0$, $\chi=0.1$,
$\kappa=8.0$, and $t_{m}=0.5 \Gamma^{-1}_{d}$.   }
\end{figure}

\begin{figure}
  \includegraphics[scale=0.65]  {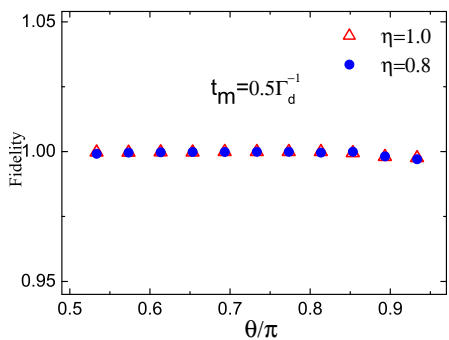}\\
  \caption{
Alternative plot of the result in Fig.\ 2,
via the fidelity of the estimated
state $\rho$ with respect to the ``true" one,
$\varrho_i=|\psi_i\ra\la\psi_i|$,
using $F={\rm Tr}(\varrho_i \rho)$.  }
\end{figure}

\begin{figure}
  \includegraphics[scale=0.7] {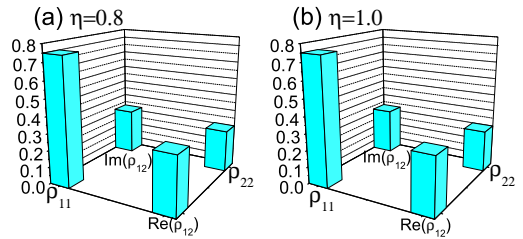}\\
  \caption{
Tomographic plot of the estimated state from the result
in Fig.\ 2 (an example using post-selection with $\theta=0.65\pi$).
Detailed numerics:
The ``true" (unknown) state was set as
$\varrho_{i,11}=0.75$ and $\varrho_{i,12}=0.34+0.265i$.
The state estimated from ideal measurements ($\eta=1$) is
$\rho_{11}=0.75095$ and $\rho_{12}=0.33218+0.27691i$,
while the result from $\eta=0.8$ reads
$\rho_{11}=0.74891$ and $\rho_{12}=0.3356+0.27461i$.   }
\end{figure}

% ==========================================
% Fig.2+Fig.3+Fig.4: \\
%  \\

We now turn to an important issue related to the
state tomography under present investigation.
That is, this scheme is free from the
{\it efficiency} of the quantum weak measurement.
This unexpected feature is rooted in a finding
in our previous study \cite{Li15},
where the weak values of qubit measurements
were found free from the quantum efficiency of the measurements.
Note that, in sharp contrast with this,
state tracking by continuous weak measurement
and quantum feedback control, would essentially
depend on the efficiency of the quantum measurements.
Non-ideal measurements will degrade
the fidelity of the controlled target state,
or completely lose all the state information.
This drawback is actually the main obstacle
of quantum feedback control
in the circuit QED systems \cite{Sid12}.

Within the Bayesian formalism,
we simply account for the measurement
inefficiency by inserting a decoherence factor
$e^{-(1-\eta)\Gamma_m t_m}$
into the off-diagonal elements of the qubit state.
This treatment has qualitatively included the consequences
of such as the amplifier's noise in the homodyne measurement
and the loss of measuring photons.
Accordingly, in running the effective quantum trajectory equation
\cite{Li15,Gam08}, we reduce, simultaneously,
the rates $\Gamma_{ci}$ and $\Gamma_{ba}$ by a factor ``$1-\eta$".

In Fig.\ 2 we compare the AAV WV extracted from
the ideal measurement (red triangles, with $\eta=1$)
with the one under efficiency $\eta=0.8$ (blue dots),
while plotting both against the ``true" result (solid curve).
Indeed, we find all the three results in perfect agreement.
In Fig.\ 3 we further display the fidelity of the estimated
state $\rho$ with respect to the ``true" one,
$\varrho_i=|\psi_i\ra\la\psi_i|$,
using the fidelity definition $F={\rm Tr}(\varrho_i \rho)$,
while in Fig.\ 4 we characterize, for a specific example,
the full state (diagonal and off-diagonal elements
of the density matrix) in terms of the usual means
of quantum state tomography.
Through these plots, we illustrate that, indeed,
the direct (weak value associated) scheme of quantum state
tomography is free from the efficiency of quantum measurement.

% Fig.5: Beyond bad-cavity and weak response limits:
%%
Now let us consider the situation
beyond the bad-cavity and weak response limits,
and illustrate how to reliably extract the AAV WV
and determine the qubit state.
We set $\kappa=2$ and remain all the other parameters
the same as used in Figs.\ 1-4.
In this case, if we {\it improperly} use
\Eq{cQED-WV-3} with all the rates and the ac-Stark shift
determined by the steady-state cavity fields,
as indicated by the blue dots in Fig.\ 5, the extracted
AAV WV will suffer serious error from the ``true" result.
However, instead, if we combine \Eq{cQED-WV-3} with
the factors given by \Eq{t-factor},
satisfactory results can be obtained,
as shown in Fig.\ 5 by the red dots.
This ensures that the {\it direct} scheme of state tomography
can be applied beyond the bad-cavity
and weak response limits, if one properly applies
\Eqs{cQED-WV-3} and (\ref{t-factor}).

\begin{figure}
  \includegraphics[scale=0.65] {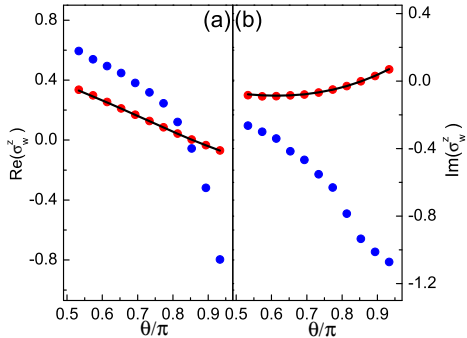}\\
  \caption{
AAV WVs extracted from measurements
beyond the bad-cavity and weak response limits.
Red dots: results extracted {\it correctly} using \Eq{cQED-WV-3}
together with the factors \Eq{t-factor}.
Blue dots: results extracted {\it improperly}
using \Eq{cQED-WV-3} under steady state of the cavity fields,
as done in the bad-cavity and weak response limits.
The ``true" results are displayed by the solid curves.
Parameters:
$\Delta_r=0$, $\epsilon_{m}=1.0$, $\chi=0.1$,
$\kappa=2.0$, and $t_{m}=0.5 \Gamma^{-1}_{d}$.  }
\end{figure}

%\clearpage

\section{Summary and Discussion}

{\flushleft
We have presented a new scheme for qubit state tomography}
in the superconducting circuit-QED system, based on
weak measurements and the associated quantum Bayesian approach.
The Bayesian approach allows us to derive a compact expression
for the PPS average, which encodes
the full information of the AAV WV
and makes the participation of
its real and imaginary parts tunable
by modulating the LO phase of the homodyne measurement.
For the first time, we also obtained analytic expression
for the PPS average {\it beyond}
the bad-cavity and weak-response limits,
and demonstrated how to determine the qubit state in this regime.

We may stress that, in order to reduce the measurement disturbance
on the measured state, ``weakness" of measurement
is usually required to the weak-value-based {\it direct} scheme.
However, differing from state tracking and feedback control,
the direct state tomography is free from the {\it efficiency}
of the quantum weak measurement.
This feature is out of simple expectation,
since the non-ideality of measurement will affect
state inferring conditioned on the measurement results,
and thus affect the success probability of post-selection.
The key point is that
the PPS average is free from the efficiency of measurement.
This {\it efficiency-free} feature can greatly benefit
the implementation of the proposed scheme in experiments.

It would be of interest to explore the direct scheme
of state tomography for more complicated states,
e.g., entangled state of multiple qubits,
and nontrivial quantum state of cavity fields.
We may leave such sort of problems for future investigations.

% =====================================================
\vspace{0.2cm}
{\flushleft \it Acknowledgments.}---
This work was supported by the NNSF of China
under grants No.\ 91321106 \& 210100152,
the State ``973" Project under grant No.\ 2012CB932704,
the Beijing NSF under grant No.\ 1164014,
and the Fundamental Research Funds for the Central Universities.

%% \clearpage

%\end{CJK*}
\end{document}